\documentclass[aps,prd,showkeys,showpacs,preprint,groupedaddress]{revtex4}
\usepackage{graphicx}

\begin{document}

\title{Quantum Electrodynamics is free from the Einstein-Podolsky-Rosen paradox}

\author{Daniele Tommasini}

\affiliation{Departamento de F\'\i sica Aplicada, \'Area de
F\'\i sica Te\'orica, Universidad de Vigo, 32004 Ourense, Spain}

\email[]{daniele@uvigo.es}

\date{October 9, 2001; Revised Version: December 9, 2001}

\begin{abstract}
I show that Quantum Electrodynamics (QED) predicts a sort of
uncertainty principle on the number of the ``soft photons" that
can be produced in coincidence with the particles that are
observed in any EPR experiment. This result is argued to be
sufficient to remove the original EPR paradox. A signature of this
soft-photons solution of the EPR paradox would be the observation
of apparent symmetry violation in single events. On the other
hand, in the case of the EPR experiments that have actually been
realized, the QED correlations are argued to be very close to
those calculated by the previous, incomplete treatment, which
showed a good agreement with the data. Finally, the usual
interpretation of the correlations themselves as a real sign of
nonlocality is also criticized.
\end{abstract}



\eprint{quant-ph/0110059}

\maketitle

\section{\label{sec:1}Introduction}

In their famous 1935 paper \cite{EPR}, Einstein, Podolsky and
Rosen (EPR) pointed out that the Quantum Mechanics probabilistic
description of Nature apparently leads to some mysterious action
at a distance. They eventually deduced that the Quantum Theory
itself was necessarily incomplete. This suggested the need for
some Hidden Variables, allowing for a causal local, deterministic
description, but such an hypothesis can hardly agree with the
results of a series of experiments carried out in the last several
years \cite{Bell,Aspect}.

It is now commonly believed that local realism is violated by
quantum effects even in the relativistic case \cite{Aharonov}.
This is so paradoxical, that some authors still suggest the need
for a ``more consistent" theory beyond the present day
Relativistic Quantum Mechanics, and argue that a conclusive
experimental proof against Hidden Variables is lacking
\cite{Santos}.

Actually, it turns out that no new physics is needed. In fact,
there is already a very elegant theory, which describes with
extreme accuracy all the known phenomena involving the
electromagnetic interaction \footnote{Moreover, it is a part of
the Glashow-Weinberg-Salam Standard Model, that describes all
known interactions except gravity.}. It is Quantum Electrodynamics
(QED) \cite{Tomonaga,WeinbookI}. Since it is by construction
relativistic, and based on the local U(1) gauge principle, it is
not surprising that it is protected against the EPR paradox, as I
show in the present paper.

\section{\label{sec:2}The EPR paradox}

Let me consider an ideal EPR experiment \cite{EPR,BJ}. Two
particles, e.g.˜ two photons as in the actual experiments
\cite{Aspect}, are emitted by a source and travel in opposite
directions \footnote{It is easy to generalize the present
discussion to the case of three or more particles.}. Far apart,
some conserved observable, such as energy, momentum, or a
component of the angular momentum (spin, helicity or
polarization), is measured on both of them.

According to the usual interpretation, the measurement carried on
one of the two subsystems (call it A) reduces it to an eigenstate
of the measured observable, whose conservation immediately forces
also the second particle (call it B) to ``collapse" into the
corresponding ``entangled" eigenstate. Since before the
experiments the two particles are not prepared in an eigenstate of
the measured quantity (see also point 2 few lines below), it seems
that the observation on A implies an instantaneous change of the
state of the distant particle B. The observed quantity would then
get ``an element of physical reality", according to the original
definition: {\it ``If, without in any way disturbing a system we
can predict with certainty (i.e.˜ with probability equal to unity)
the value of a physical quantity, then there is an element of
physical reality corresponding to this physical quantity"}
\cite{EPR} (the italics and the parenthesis also belong to the
original paper).

According to Einstein and collaborators, the problem of Quantum
Mechanics is that: 1) the considered observable gets a {\it
definite} value on B, after a measurement on the distant particle
A, {\it and this occurs with certainty,} so that the observable
gets a ``physical reality" on B; 2) such a physical reality
depends on the actual measurement that is done on A (for instance,
if instead of measuring the component $J_z$ of the angular
momentum we decided to measure an observable incompatible with it,
such as $J_x$, then the state of the distant particle B would
correspond to a {\it different} physical reality) \cite{EPR}. Such
a situation is also called a violation of ``local realism".

This is the original EPR paradox of Quantum Mechanics. Only much
later, mainly due to the work of Bell \cite{Bell}, it was
reformulated in terms of the correlations (see also sections
\ref{sec:4} and \ref{sec:5}), in order to work out predictions
that could be tested experimentally. Up to now, the two
formulations were thought to be equivalent, but in section
\ref{sec:4} we shall see that this is not the case. For the
moment, let us notice that, according to Einstein and
collaborators, the existence of an action at a distance working
with perfect efficiency would imply a much harder incompatibility
with Special Relativity, as compared with a possible statistical
nonlocality as shown in the correlations. I will come back later
to the problem of the correlations, and concentrate for the moment
on the original EPR paradox.

It is worth mentioning that Einstein and collaborators concluded
that the solution to the paradox should necessarily imply that
``the wave function does not provide a complete description of the
physical reality" \cite{EPR}. For the wave function, they meant
that associated to the two particle system, describing A and B and
nothing else. We shall see that they were right, even though they
would perhaps not expect that the solution was to be found in the
modern version of the Quantum Theory itself. In fact, the complete
description in QED is not limited to the ``entangled" state of A
and B, since it allows also for the presence of an arbitrary
number of ``soft photons".

\section{\label{sec:3}The uncertainty in the number of soft photons}

The EPR paradox, as described above, is originated by the
assumption of a two particle state, which is incorrect in
Relativistic Quantum Mechanics. As we shall see, states involving
two or more particles are not ``stable" in QED. There are no
entangled ``stationary" states! In other (more correct) words,
{\it additional real particles can be created in coincidence with
A and B.} Which additional species can appear depends on the
available energy. Since massless particles can have arbitrarily
low energy, the possible presence of real ``soft photons" (i.e.˜
very low energy photons) should always be taken into account in
the theoretical treatment. In any case, since soft photons usually
escape detection (or they are not looked for), no event can be
guaranteed with absolute certainty to involve only the two
observed ``hard" particles.

Moreover, soft photons can also be created due to the interaction
of both A and B with the measuring apparatus. Even though the
latter effect will not be used in the rest of the present work, in
this section I will mention it since it can be interesting for the
Theory of Measurement (for instance, due to possible creation of
soft photons during the measurement, the ideal measurement that is
used to define the eigenstates of the observables should be
considered as a mere approximation).

\begin{figure}
\includegraphics{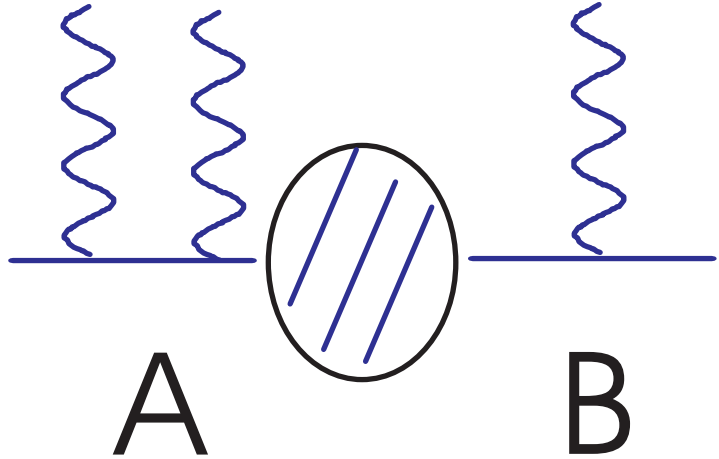}
\caption{\label{fig1} Feynman diagram describing the production of
an EPR pair of charged spin 1/2 particles, A and B, in coincidence
with three soft photons. The dashed blob represents the part
depending on the particular basic process and the initial
particles that are considered.}
\end{figure}

\begin{figure}
\includegraphics{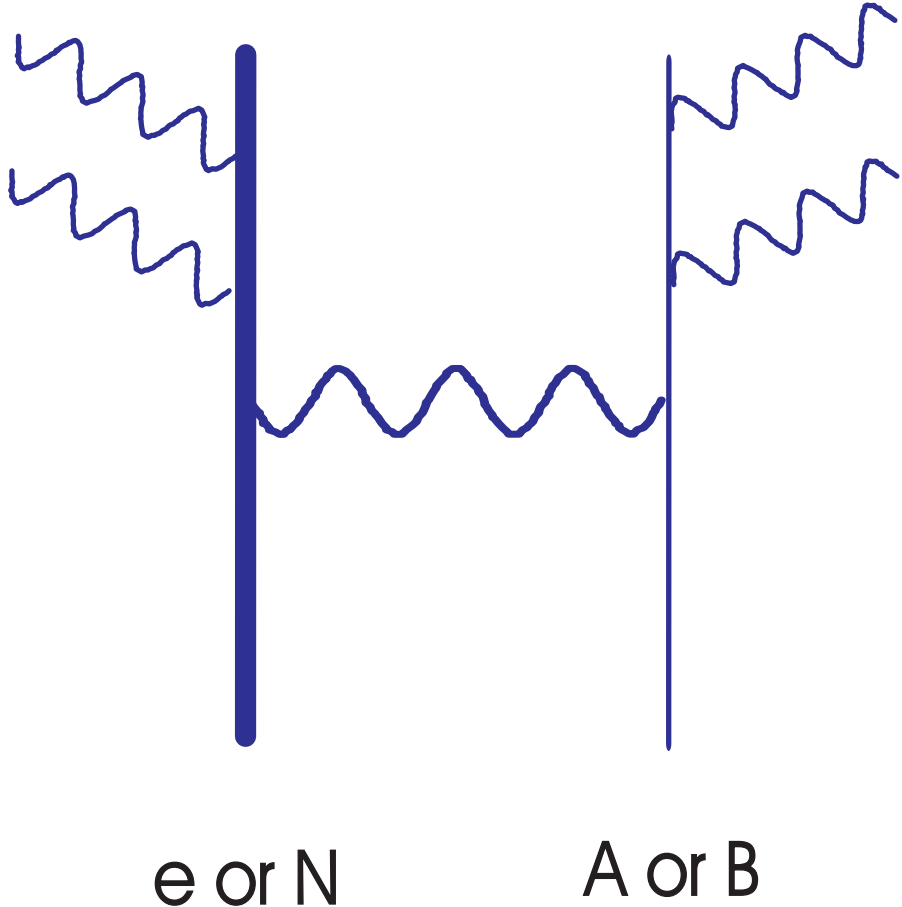}
\caption{\label{fig2} Feynman diagram describing the creation of
four soft photons due to the interaction of any one of the EPR
spin 1/2 particles A,B, with an electron e or a nucleus N of the
measuring apparatus.}
\end{figure}

\begin{figure}
\includegraphics{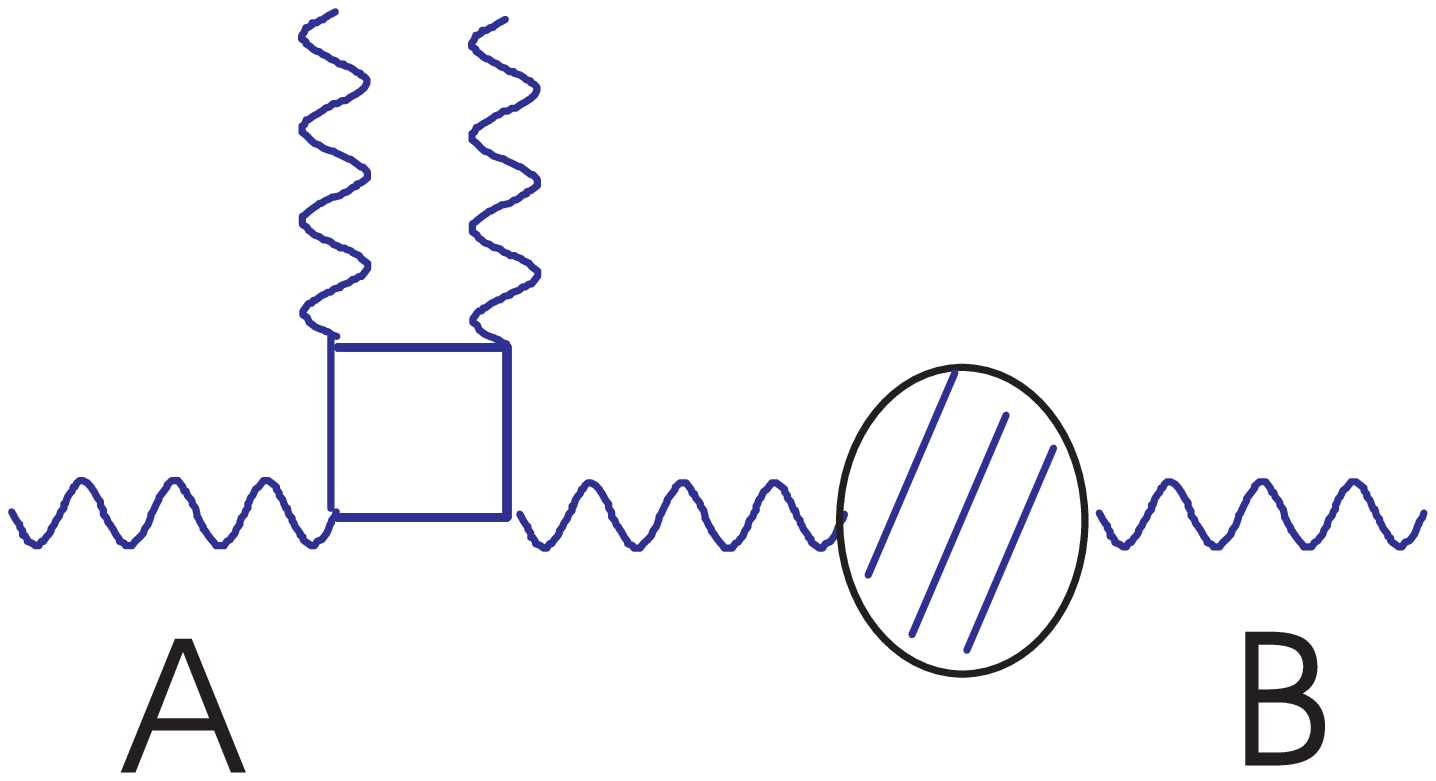}
\caption{\label{fig3} Feynman diagram describing the production of
an EPR pair of photons, A and B, in coincidence with two soft
photons. The dashed blob represents the part depending on the
particular basic process and the initial particles that are
considered.}
\end{figure}

\begin{figure}
\includegraphics{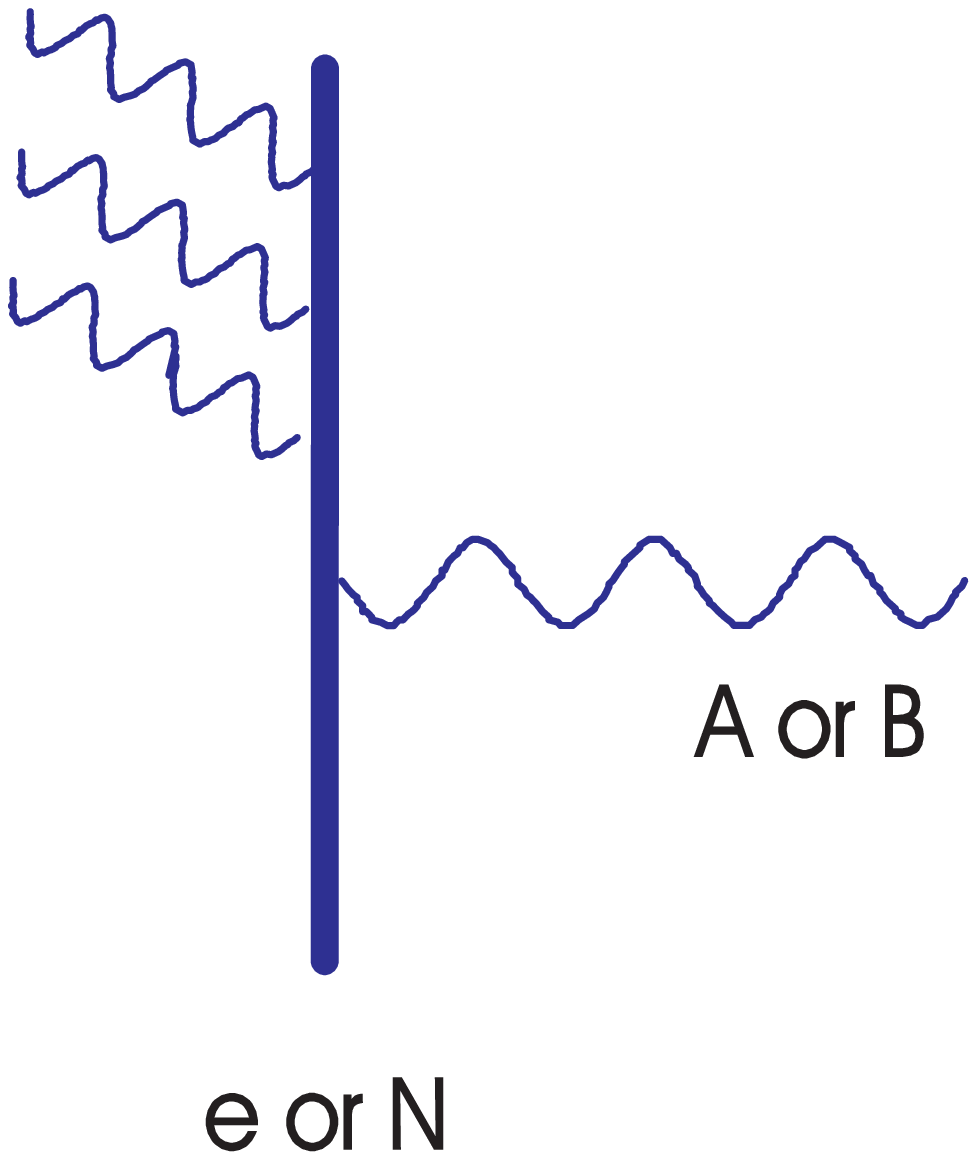}
\caption{\label{fig4} Feynman diagram describing the production of
three photons (not all of them are necessarily ``soft") due to the
interaction of any one of the EPR photons A,B, with an electron e
or a nucleus N of the measuring apparatus.}
\end{figure}

According to the previous discussion, there are two sources of
indetermination on the number of {\it real} particles in an EPR
experiment: at the production process, or at the measuring
apparatus. I will prove this statement using QED perturbation
theory (i.e.˜ Feynman diagrams). For simplicity, I will only
discuss two kinds of EPR experiments: i) those involving two
charged spin 1/2 particles; and ii) those involving two photons.
In both cases, I will give explicit examples predicting the
creation of an arbitrary number of soft photons.

i) In Fig.˜ \ref{fig1}, I have drawn a tree-level diagram where
the blob represents the particular elementary process that
produces particles A and B.
Even without specifying that part of the diagram (involving some
``initial" particles), we see that an arbitrary number of real
soft photons (three in the particular case of the figure) can be
attached to each of the external fermion legs. This is a well
known effect in QED \cite{WeinbookI}. Fig.˜ \ref{fig2} shows a
diagram describing the interaction of any one of our two particles
with a charged particle belonging to the measuring apparatus,
through the exchange of a virtual photon.
Here, an arbitrary number of {\it real} soft photon lines can be
attached to both the electron under measurement and the charged
particle belonging to the experimental device.

ii) The two photon case, which corresponds e.g.˜ to Aspect et al.˜
experiment \cite{Aspect}, seems to be a bit more complicated from
a theoretical point of view. Since no three photon vertex exist,
we have to look for one loop effects. In Fig.˜ \ref{fig3}, I show
a ``box" diagram for the production of two real soft photons
\footnote{This box diagram has been studied in different contexts,
e.g.˜ in the theory of two photon scattering.}.
The virtual particle in the loop can be any charged fermion
(electron, muon, tau, quarks). On the other hand, the interaction
with the measuring apparatus can proceed through diagrams such as
the tree level one of Fig.˜ \ref{fig4}.
Here, soft photons can be attached in an arbitrary number to the
line of the electron or nucleon belonging to the experimental
detector \footnote{Actually, in the common experimental situation
\cite{Aspect} the ``hard" photon energy is comparable with the
electron binding energies in matter, and QED perturbation theory
breaks down \cite{WeinbookI}. But even if it will be hard to {\it
compute} its rate, soft photon emission during the measurement
will still be possible.}.

It is clear that these considerations can be generalized: an
arbitrary and unknown number of soft photons can always be created
in any experiment, in any step that involves an interaction.

In the following discussion, we will be interested in particular
in the soft photons that are created in coincidence with the two
(or more) particles observed in an EPR experiment, as described by
diagrams such as those in Figs.˜ 1 and 3. We are now able to
understand how QED is protected from the EPR paradox.

\section{\label{sec:4}The QED solution to the EPR paradox}

After the measurement of the conserved observable, say a component
of the angular momentum, is performed on particle A, we cannot
have any idea about how many soft photons are there around. QED
predicts that the conservation law holds for the set including
particles A and B together with all the soft photons that are
created with them, through diagrams such as those in Figs.˜ 1 or
3. Therefore, we could only say that the subsystem consisting of
particle B and all the soft photons that have been produced in
coincidence with them will get a definite value of the angular
momentum, determined by the result obtained on A. This means that
{\it the measurement on A does not allow for a certain prediction
of the value of the considered conserved quantity on B}.
Therefore, the (component of the) angular momentum {\it will not
be given a ``physical reality"} on B after the measurement on A.
According to the discussion of section \ref{sec:2}, this is
sufficient to save the theory from the original EPR paradox.

It is important to recall that there is no possibility to control
completely the uncertainty on the number of soft photons in a
single event. In other words, QED is even less deterministic than
Nonrelativistic Quantum Mechanics, due to this underlying sort of
Uncertainty Principle on the Number of Particles. In fact, the
only predictions that it allows are on probabilities and average
values. This greater indetermination protects the theory from the
EPR paradox. In other words, it seems that, to remove the paradox,
one has to choose between the most extreme possibilities:
determinism (hidden variables, the favorite option for Einstein
and collaborators), or complete lack of determinism for the single
processes (QED, the dice of God) \footnote{However, QED is also a
deterministic theory, since the field equations are deterministic.
The correlation functions themselves are determined by the initial
conditions and the interaction lagrangian. This is an important
point, whose possible consequences will be discussed below.}.

Let me come back to our EPR experiments, and notice that the
conservation laws, including energy and momentum, are not expected
to hold strictly for the two particle (sub)system, A and B. A
general single event will show apparent symmetry violations,
except when by chance no soft photon is created. In particular,
any violation of a discrete variable such as angular momentum is
important, since it is a multiple of $\hbar$. These considerations
suggest that a possible signature of the theoretical explanation I
am proposing would be the observation of an apparent symmetry
violation event in an EPR experiment. This would confirm the
presence of unobserved soft photons, resulting in a further
triumph of QED.

Notice that in the actual EPR experiments \cite{Aspect} the
correlations between the polarizations of the ``hard" photons (A
and B) are evaluated. Such correlations are statistical averages
over the results for the different single events.  The data agree
with the prediction of a Quantum Mechanics that did not take into
account the soft photons, and are incompatible with the
predictions of Hidden Variables theories, that were also
considered to be the only possible locally realistic theories.
This fact was interpreted as a proof that Nature is EPR
paradoxical. However, such a conclusion is not correct. As we will
see in the next section, in the case of the EPR experiments that
have been performed up to now the QED prediction for the
correlations is very close to that obtained in Quantum Mechanics
by ignoring the soft photons, so that it can still agree with the
data within the experimental errors. However, even a very small
probability for soft photons creation is sufficient to forbid any
{\it certain} prediction for the measurement on B as a consequence
of the measurement on A, and this is enough to remove the original
EPR paradox, as we have seen. This implies that the experimental
study of the correlations {\it cannot} be used to decide about the
original EPR paradox; it can be used only to disprove the Hidden
Variable solution. In other words, the equivalence between the
original EPR paradox and its version in terms of the correlations,
as worked out by Bell, would hold only if the theoretical
correlations were strictly maximal (i.e.˜ if the soft photons did
not exist). On the other hand, the observation of a single event
showing an apparent violation of the considered conservation law
would be an evidence in favor of the present solution for the EPR
paradox, since it would confirm that no certain prediction can be
made on the state of B after measuring on A.

Notice also that this solution to the EPR paradox is based on two
points: the existence of massless particles, the photons; and of
the fermion-photon vertex, that allows photons to be created in
any external line of the relevant Feynman diagrams. But it is well
known that both the electromagnetic vertex and the masslessness of
the photon are the direct consequences of the local gauge
symmetry. Not only the local symmetry defines the theory, but it
also protects against the EPR paradox and the violation of local
realism.

\section{\label{sec:5}The QED prediction for the correlations}

Even though we have found that the study of the correlations is
not relevant for the original EPR paradox, we have to check that
the QED predictions still agree with the experiments. The
calculation can be done by using the methods discussed in Ref.˜
\cite{WeinbookI}, and the result depends on the actual selectivity
in energy and momentum of the experimental setting. Here, I will
just provide a rough argument to show that the correlations are
usually not expected to be seriously modified by the soft photons
creation.

For simplicity, I will consider an ideal EPR experiment involving
two charged spin 1/2 particles created after the decay of a zero
spin system (let us forget here the difficulty in the measurement
of the spin of the charged particles). In this case, the relevant
correlation functions are the average values of the products of
the components $S_{\vec u}(A)$ and $S_{\vec v}(B)$ of the spins of
the two particles along arbitrary axes $\vec u$ and $\vec v$
\cite{BJ}. For instance, let $\vec u=\vec v$, chosen to be the z
axis. If we do not take into account the soft photons, according
to Quantum Mechanics the two particles must have opposite spins in
order to conserve the total angular momentum. Then we get
\begin{equation}
<S_z(A)S_z(B)>=-\frac{\hbar^2}{4}. \label{Bellnonrel}
\end{equation}
Notice that this is the maximal correlation (in absolute value)
that can be achieved for two observables whose eigenvalues are
$\pm\hbar/2$. (If the spins were completely independent,
$<S_z(A)S_z(B)>=<S_z(A)><S_z(B)>=0 $.)

In general, allowing for the soft photons creation through
diagrams similar to that of Fig.˜ 1, the correlation will be
smaller than maximal. Now, as shown in Chapter 13 of Ref.˜
\cite{WeinbookI}, in the limit where the energy of the soft
photons is neglected the helicities of the two fermion will remain
opposite. Therefore, the correction to Eq.˜ (\ref{Bellnonrel}) due
to diagrams such as that of Fig.˜ 1 is suppressed by powers of
$\left(\frac{E_{soft}}{E}\right)^2$, where $E$ is the ``hard"
fermions energy and $E_{soft}$ is the typical soft photons energy
(essentially, it is the ``infrared cutoff" introduced in Ref.˜
\cite{WeinbookI}). This parameter depends on the experimental
settings, but it can be made small by increasing the energy
selectivity in the observation of particles A and B. Moreover, in
a ``selective enough" experimental setting, the two particles will
be detected in opposite directions with small angular
indetermination, then the total transversal momenta of the soft
photons will have a limited phase space available, and this will
result in a further suppression of the corresponding diagrams.
Diagrams involving an increasing number of soft photons will also
be suppressed by the corresponding powers of the fine structure
constant $\alpha\simeq1/137$. For all these reasons, in the usual
EPR experiments we expect that the correlation will be close to
that computed using the ``entanglement" theory, and the agreement
with the data will not be spoiled within the experimental errors.

However, as we have discussed, even a very small probability for
soft photons creation is sufficient to save the theory from the
original EPR paradox, since it prevents the possibility of a
certain prediction on the single event. For instance, in event
involving a single photon travelling close to the direction of the
two particles A and B, the two fermion helicities will most
probably be found parallel rather than antiparallel, in order to
cancel the $\pm\hbar$ helicity of the photon.

A similar result can be found in the case of the actual EPR
photons experiments. In fact, the probability due to the relevant
diagram (Fig.˜ 3) is suppressed by four powers of the fine
structure constant, by the reduced phase space and by the electron
propagators in the loop (the mass of the electron is large as
compared with the typical energy of the ``hard" photons involved
in the experiment, which are typically in the eV range).
Therefore, the prediction obtained in QED, taking into account the
soft photons, is expected to be very close to that of the previous
approach, and will still agree with the present experimental
results.

\section{\label{sec:6}The problem of the apparent nonlocality of the QED
correlations}

In the QED correlations that we have discussed above, the
conservation laws hold for the set of particles A and B and all
the possible soft photons. Everything goes as though there were a
secret agreement amongst all the distant particles that can appear
in a single event (including those that are not observed). This
fact is often interpreted as a sign of some ``quantum
nonlocality". Before discussing this point, I would like to
remember that this problem of the EPR correlations {\it is not}
the EPR paradox, that is removed by taking into account the
uncertainty on the soft photons (see sections \ref{sec:2} and
\ref{sec:4} above). On the other hand, the correlations do not
imply any direct violation of Special Relativity, since they are
merely a statistical property (at least, this was the point of
view of Einstein and collaborators, see also few lines below).

In principle, it could be hoped that such an apparent nonlocality
of the correlations could be used to save some supposed
applications of the EPR paradox, such as teleportation
\cite{telepo}, that might thus be interpreted as intrinsically
statistical processes. For instance, if teleportation is realized
using (``hard") photons in the eV range, the probability for soft
photons creation is very small, as we have discussed above, and
the existing theory could be thought to be a good approximation.
However, I think that a deep study of the measurement problem is
needed to prove whether such an interpretation can be correct.
This point is of extreme importance and urgency, since
teleportation is presently used as a base for Quantum Information
Theory and Quantum Computing.

Here, I will provide a possible qualitative argument against the
nonlocality interpretation of the correlations, without pretending
it to be definitive, in order to stimulate a debate on such an
urgent problem. In fact, in QED the correlations are obtained from
a covariant Lagrangian density that only involves {\it local}
interactions; they are causal and the prediction for them is {\it
deterministic} \cite{WeinbookI}. The fact that the Feynman
diagrams of the kind of Figs.˜ 1 and 3 imply a conservation of
energy, momentum, angular momentum, etc., amongst the external
legs, is a causal, deterministic consequence of the initial ``in"
state and the local interaction that occurs at the production
point. Therefore, I think that it is not correct to interpret the
global conservation law as a result of an ``instantaneous
agreement" occurring at the moment of the measurement, since the
correlations showing such a global conservation are calculated as
a deterministic result of the evolution from the common origin of
the particles A, B and the soft photons. In other words, the
global conservation is merely a causal consequence of the local
conservation law. No mysterious action at a distance is then
working. The real ``quantum mystery" is the wave-particle duality,
with the localization of the particles in the single events. But
the amplitude of probability is a wave, whose evolution respects
causality {\it and} locality. I think that for this reason
Einstein and collaborators in Ref.˜ \cite{EPR} were not concerned
with the (nonmaximal) probabilities or correlations, and they were
so careful in defining the paradox as a problem occurring if A and
B were perfectly ``entangled" (``with certainty, i.e.˜ with
probability equal to one", as they say explicitly). Therefore,
since we have found that this original paradox (i.e.˜ the
``perfect entanglement") is removed by the soft photons mechanism
(even with a very small probability for them to be created), I
think that also the nonlocality interpretation of the correlations
is undermined. At least, such a ``sort of nonlocality", that
(roughly speaking) originates from the causal and local
propagation of a ``wave" from the production point, does not
correspond to any mysterious action at a distance, and cannot be
used e.g.˜ as a base for teleportation.

\section{\label{sec:7}Conclusions}

To conclude, I have shown that QED is protected from the original
EPR paradox by the local gauge symmetry. This corresponds to the
fact that it allows for the creation of an arbitrary number of
soft photons in coincidence with the observed particles in an EPR
experiment. This mechanism would be confirmed by the experimental
observation of an apparent symmetry violation in a single event.
On the other hand, the correlations are expected to be smaller
than those calculated by ignoring the soft photons, but in the
case of the actual EPR experiments that have been realized up to
now the correction is expected to be very small, so that the
agreement of the Quantum Theory with the present data is not
spoiled. Such correlations are usually thought to be by themselves
a sign of a ``quantum nonlocality". Although here I have already
presented an argument against such an interpretation, this problem
deserves further research, which is particularly urgent in order
to decide about the actual viability of several supposed
applications of the Quantum Theory that were based on the EPR
paradox.

\begin{acknowledgments}
It is a pleasure to thank Humberto Michinel for stimulating
discussions and help, and Esther P\'erez, Rafael A. Porto, Uwe
Trittmann and Ruth Garc\'\i a Fern\'andez for useful comments.
\end{acknowledgments}

\bibliography{QEDEPR_N}

\begin{thebibliography}{9}
\expandafter\ifx\csname natexlab\endcsname\relax\def\natexlab#1{#1}\fi
\expandafter\ifx\csname bibnamefont\endcsname\relax
  \def\bibnamefont#1{#1}\fi
\expandafter\ifx\csname bibfnamefont\endcsname\relax
  \def\bibfnamefont#1{#1}\fi
\expandafter\ifx\csname citenamefont\endcsname\relax
  \def\citenamefont#1{#1}\fi
\expandafter\ifx\csname url\endcsname\relax
  \def\url#1{\texttt{#1}}\fi
\expandafter\ifx\csname urlprefix\endcsname\relax\def\urlprefix{URL }\fi
\providecommand{\bibinfo}[2]{#2}
\providecommand{\eprint}[2][]{\url{#2}}

\bibitem[{\citenamefont{Einstein et~al.}(1935)\citenamefont{Einstein, Podolsky,
  and Rosen}}]{EPR}
\bibinfo{author}{\bibfnamefont{A.}~\bibnamefont{Einstein}},
  \bibinfo{author}{\bibfnamefont{B.}~\bibnamefont{Podolsky}}, \bibnamefont{and}
  \bibinfo{author}{\bibfnamefont{N.}~\bibnamefont{Rosen}},
  \bibinfo{journal}{Phys.\ Rev.} \textbf{\bibinfo{volume}{47}},
  \bibinfo{pages}{777} (\bibinfo{year}{1935}).

\bibitem[{\citenamefont{Bell}(1964)}]{Bell}
\bibinfo{author}{\bibfnamefont{J.}~\bibnamefont{Bell}},
  \bibinfo{journal}{Physics} \textbf{\bibinfo{volume}{1}}, \bibinfo{pages}{195}
  (\bibinfo{year}{1964}).

\bibitem[{\citenamefont{Aspect et~al.}(1981)\citenamefont{Aspect, Grangier, and
  Roger}}]{Aspect}
\bibinfo{author}{\bibfnamefont{A.}~\bibnamefont{Aspect}},
  \bibinfo{author}{\bibfnamefont{P.}~\bibnamefont{Grangier}}, \bibnamefont{and}
  \bibinfo{author}{\bibfnamefont{G.}~\bibnamefont{Roger}},
  \bibinfo{journal}{Phys.\ Rev.\ Lett.} \textbf{\bibinfo{volume}{47}},
  \bibinfo{pages}{460} (\bibinfo{year}{1981});
\bibinfo{author}{\bibfnamefont{A.}~\bibnamefont{Zeilinger}},
  \bibinfo{journal}{Rev.\ Mod.\ Phys.} \textbf{\bibinfo{volume}{71}},
  \bibinfo{pages}{S288} (\bibinfo{year}{1999}).

\bibitem[{\citenamefont{Aharonov and Albert}(1981)}]{Aharonov}
\bibinfo{author}{\bibfnamefont{Y.}~\bibnamefont{Aharonov}} \bibnamefont{and}
  \bibinfo{author}{\bibfnamefont{D.}~\bibnamefont{Albert}},
  \bibinfo{journal}{Phys.\ Rev. D} \textbf{\bibinfo{volume}{24}},
  \bibinfo{pages}{359} (\bibinfo{year}{1981});
\bibinfo{author}{\bibfnamefont{G.}~\bibnamefont{Ghirardi}},
  \bibinfo{author}{\bibfnamefont{R.}~\bibnamefont{Grassi}}, \bibnamefont{and}
  \bibinfo{author}{\bibfnamefont{P.}~\bibnamefont{Pearle}},
  \bibinfo{journal}{Found.\ Phys.} \textbf{\bibinfo{volume}{20}},
  \bibinfo{pages}{1271} (\bibinfo{year}{1990});
\bibinfo{author}{\bibfnamefont{L.}~\bibnamefont{Hardy}},
  \bibinfo{journal}{Phys.\ Rev.\ Lett.} \textbf{\bibinfo{volume}{68}},
  \bibinfo{pages}{2981} (\bibinfo{year}{1992});
\bibinfo{author}{\bibfnamefont{I.}~\bibnamefont{Percival}},
  \bibinfo{journal}{Phys.\ Lett. A} \textbf{\bibinfo{volume}{244}},
  \bibinfo{pages}{495} (\bibinfo{year}{1998}).

\bibitem[{\citenamefont{Santos}(1992)}]{Santos}
\bibinfo{author}{\bibfnamefont{E.}~\bibnamefont{Santos}},
  \bibinfo{journal}{Phys.\ Rev.\ A} \textbf{\bibinfo{volume}{46}},
  \bibinfo{pages}{3646} (\bibinfo{year}{1992});
\bibinfo{author}{\bibfnamefont{L.~D.} \bibnamefont{Caro}} \bibnamefont{and}
  \bibinfo{author}{\bibfnamefont{A.}~\bibnamefont{Garuccio}},
  \bibinfo{journal}{Phys.\ Rev. A} \textbf{\bibinfo{volume}{54}},
  \bibinfo{pages}{174} (\bibinfo{year}{1996});
\bibinfo{author}{\bibfnamefont{C.}~\bibnamefont{Thompson}},
  \bibinfo{journal}{Found.\ Phys.\ Lett.} \textbf{\bibinfo{volume}{9}},
  \bibinfo{pages}{357} (\bibinfo{year}{1996}).

\bibitem[{\citenamefont{Tomonaga}(1948)}]{Tomonaga}
\bibinfo{author}{\bibfnamefont{S.}~\bibnamefont{Tomonaga}},
  \bibinfo{journal}{Phys.\ Rev.} \textbf{\bibinfo{volume}{74}},
  \bibinfo{pages}{224} (\bibinfo{year}{1948});
\bibinfo{author}{\bibfnamefont{J.}~\bibnamefont{Schwinger}},
  \bibinfo{journal}{Phys.\ Rev.} \textbf{\bibinfo{volume}{73}},
  \bibinfo{pages}{416} (\bibinfo{year}{1948});
\bibinfo{author}{\bibfnamefont{R.}~\bibnamefont{Feynman}},
  \bibinfo{journal}{Rev.\ Mod.\ Phys.} \textbf{\bibinfo{volume}{20}},
  \bibinfo{pages}{367} (\bibinfo{year}{1948});
\bibinfo{author}{\bibfnamefont{F.}~\bibnamefont{Dyson}},
  \bibinfo{journal}{Phys.\ Rev.} \textbf{\bibinfo{volume}{75}},
  \bibinfo{pages}{486, 1736} (\bibinfo{year}{1949}).

\bibitem[{\citenamefont{Weinberg}(1995)}]{WeinbookI}
\bibinfo{author}{\bibfnamefont{S.}~\bibnamefont{Weinberg}},
  \emph{\bibinfo{title}{The Quantum Theory of Fields}},
  vol.~\bibinfo{volume}{I} (\bibinfo{publisher}{Cambridge University Press},
  \bibinfo{year}{1995}).

\bibitem[{\citenamefont{Bransden and Joachain}(2000)}]{BJ}
\bibinfo{author}{\bibfnamefont{B.}~\bibnamefont{Bransden}} \bibnamefont{and}
  \bibinfo{author}{\bibfnamefont{C.}~\bibnamefont{Joachain}},
  \emph{\bibinfo{title}{Quantum Mechanics}} (\bibinfo{publisher}{Prentice
  Hall}, \bibinfo{year}{2000}), \bibinfo{edition}{2nd} ed.


\bibitem[{\citenamefont{Bennet et~al.}(1993)\citenamefont{Bennet, Brassard,
  Cr\'epeau, Josza, Peres, and Wootters}}]{telepo}
\bibinfo{author}{\bibfnamefont{C.}~\bibnamefont{Bennet}},
  \bibinfo{author}{\bibfnamefont{G.}~\bibnamefont{Brassard}},
  \bibinfo{author}{\bibfnamefont{C.}~\bibnamefont{Cr\'epeau}},
  \bibinfo{author}{\bibfnamefont{R.}~\bibnamefont{Josza}},
  \bibinfo{author}{\bibfnamefont{A.}~\bibnamefont{Peres}}, \bibnamefont{and}
  \bibinfo{author}{\bibfnamefont{W.}~\bibnamefont{Wootters}},
  \bibinfo{journal}{Phys.\ Rev.\ Lett.} \textbf{\bibinfo{volume}{70}},
  \bibinfo{pages}{1895} (\bibinfo{year}{1993}).


\end{thebibliography}

\end{document}